\documentclass[fleqn,usenatbib]{mnras}

\usepackage{newtxtext,newtxmath}
\usepackage[T1]{fontenc}
\usepackage{graphicx}
\usepackage{amsmath,scalerel}
\usepackage{relsize}
\usepackage{subcaption}
\usepackage{soul}
\usepackage{booktabs}
\usepackage{comment}
\usepackage{ulem}

\DeclareRobustCommand{\VAN}[3]{#2}
\let\VANthebibliography\thebibliography
\def\thebibliography{\DeclareRobustCommand{\VAN}[3]{##3}\VANthebibliography}

\definecolor{sk}{rgb}{0.99, 0.00, 0.5}

\title[Morphology vs location in the cosmic web]{The relationship between morphology, density, and location in the cosmic web from massive to dwarf galaxies}

\author[I. Lazar et al.]{I. Lazar\thanks{E-mail: i.lazar@herts.ac.uk},$^{1}$ S. Kaviraj,$^{1}$ A. E. Watkins,$^{1}$  C. J. Conselice,$^{2}$ D. Kakkad,$^{1}$ T. M. Sedgwick,$^{1}$ G. Martin,$^{3}$ \newauthor S. Koudmani$^{1,4}$ \\
$^{1}$Centre for Astrophysics Research, Department of Physics, Astronomy and Mathematics, University of Hertfordshire, College Lane, Hatfield AL10 9AB, UK\\
$^{2}$Jodrell Bank Centre for Astrophysics, University of Manchester, Oxford Road, Manchester M13 9PL, UK\\
$^{3}$School of Physics and Astronomy, University of Nottingham, University Park, Nottingham NG7 2RD, UK\\
$^{4}$St Catharine's College, University of Cambridge, Trumpington Street, Cambridge CB2 1RL, UK\\
}

\pubyear{\the\year{}}

\begin{document}
\label{firstpage}
\pagerange{\pageref{firstpage}--\pageref{lastpage}}
\maketitle

\begin{abstract}
We study how morphology relates to environment from massive to dwarf galaxies using, for the first time, a mass-complete sample of $\sim$13,000 galaxies, in the stellar-mass and redshift ranges 10$^8$ M$_{\odot}$ < $M_{\star}$ < 10$^{11.5}$ M$_{\odot}$ and $0.2<z<0.4$, respectively. By combining \textit{\textit{HST}}-derived visual morphological classifications with local density and galaxy distances from nodes and filaments, we quantify how early-type and late-type galaxies (ETGs and LTGs) differ with respect to environment. The overall ETG fraction decreases from $\sim$65 per cent at $M_{\star}$ $\sim$ 10$^{11}$ M$_{\odot}$ to $\sim$20 per cent at $M_{\star}$ $\sim$ 10$^{8}$ M$_{\odot}$. Regardless of morphology, lower stellar mass galaxies lie further away from nodes and filaments than their more massive counterparts. While, at $M_{\star}$ $\gtrsim$ 10$^{9}$ M$_{\odot}$, ETGs reside further away from nodes and filaments than LTGs, this segregation weakens as stellar mass decreases, with ETGs and LTGs exhibiting similar locations at $M_{\star}$ $\lesssim$ 10$^{9}$ M$_{\odot}$. This diminishing difference at lower stellar mass is likely driven by the fact that filaments have a finite extent and lower mass galaxies, of all morphologies, lie further away from filament cores and are therefore confined to a smaller region of the filament itself. For high-mass galaxies (where ETGs and LTGs show strong environmental segregation), greater proximity to nodes likely inhibits coherent angular momentum acquisition, while residing closer to filament cores increases the likelihood of interactions and mergers. Both make it easier to create dispersion-dominated systems, driving the sharp rise of the ETG fraction, in the high-mass regime, close to nodes (and, to a lesser extent) filaments. Our results show that galaxy evolution is increasingly driven by internal processes as stellar mass decreases. 
\end{abstract}



\section{Introduction}
\label{sec:intro}

In the standard $\Lambda$CDM structure-formation paradigm, galaxies evolve hierarchically through successive merger and accretion events. In this model, smaller dark matter halos collapse first and subsequently merge to form progressively larger halos, while baryons cool and condense in the halo centres to form stars and create galaxies \citep[e.g.][]{White1978,Blumenthal1984}. Galaxy properties such as stellar mass, morphology, and star-formation history are therefore shaped, both by the assembly history of their host dark-matter halos and a complex interplay between gas accretion, mergers and internal feedback processes. Modern cosmological simulations broadly reproduce the observed galaxy population by coupling these hierarchical formation histories with prescriptions for star formation and feedback \citep[e.g.][]{Kauffmann1993,Croton2006,Kaviraj2017,Pillepich2019,Dubois2021,Schaye2026}.

Within this framework, environment is expected to play an important role in influencing galaxy evolution. Dense environments (such as galaxy groups and clusters) result in a variety of processes that are capable of altering galaxy properties such as morphology and colour. These include a higher frequency of mergers and interactions \citep[e.g.][]{Martin2018a}, ram-pressure stripping of cold gas by the intra-cluster medium \citep[e.g.][]{Gunn1972}, tidal perturbations and harassment \citep[e.g.][]{Moore1996,Martin2019,Jackson2021b} and strangulation, whereby a galaxy’s hot gas reservoir is removed, suppressing further star formation \citep[e.g.][]{Larson1980}. These environmentally-driven processes act in concert with internal mechanisms such as AGN and stellar feedback \citep[e.g.][]{Kaviraj2007}, to shape galaxy structure over cosmic time. 

Understanding the connection between environment and morphology is therefore of significant interest, both from an empirical point of view and as a key constraint on galaxy evolution models. At least in the massive-galaxy ($M_{\star}$ $\gtrsim$ 10$^{9.5}$ M$_{\odot}$) regime, the effect of environment on morphology is often described in terms of a `morphology-density' relation. For example, \citet{Dressler1980} demonstrate that the fraction of massive early-type galaxies (ETGs) increases with local galaxy density in clusters. This relationship appears to persist in different environments and cluster samples in the nearby Universe \citep[e.g.][]{Postman1984,Dressler1997,Goto2003,Bamford2009,Davies2025} but appears to weaken at higher redshifts \citep[e.g.][]{Smith2005,Poggianti2009,Mei2023,EuclidQ1}.\\ 

\begin{figure*}
    \centering
    \includegraphics[width=0.9\textwidth]{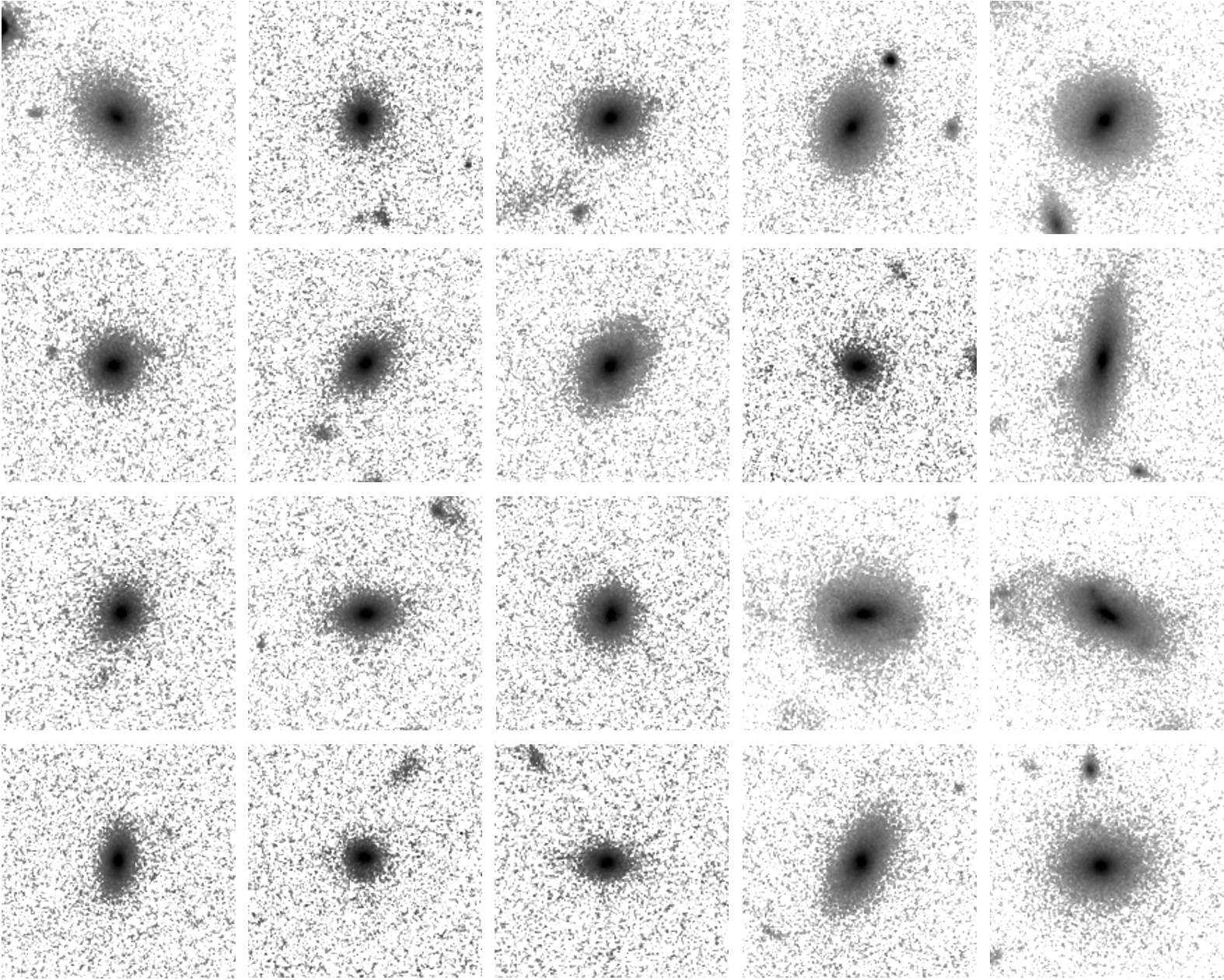}
    \caption{Examples of galaxies classified as ETGs by the visual inspection. Galaxy images are drawn from the \textit{HST} in the F814W filter. Recall that our galaxy population resides in a relatively narrow redshift range ($0.2 < z < 0.4$). All images are 5 arcseconds on a side.}
    \label{fig:etgs}
\end{figure*}

\begin{figure*}
    \centering
    \includegraphics[width=0.9\textwidth]{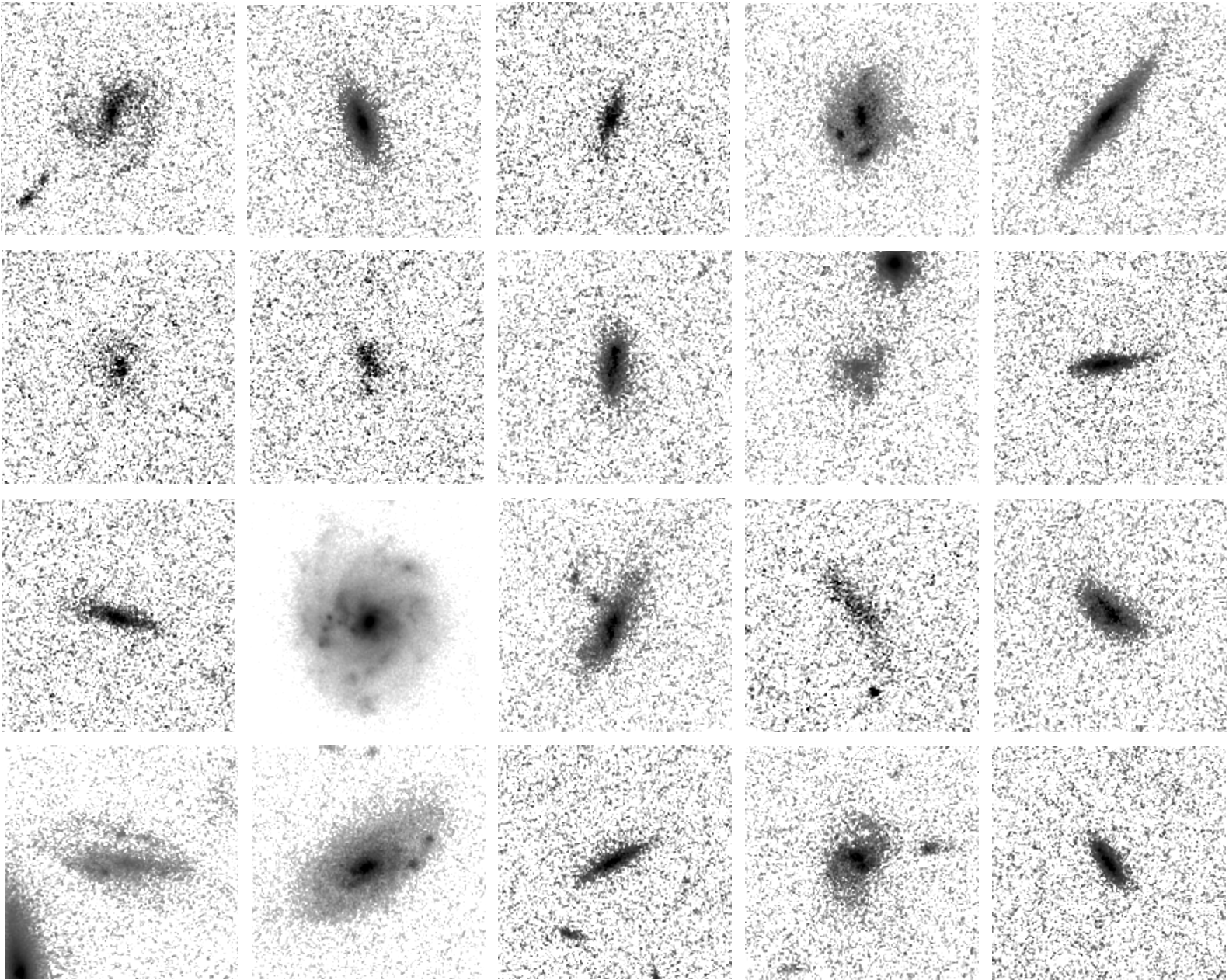}
    \caption{Examples of galaxies classified as LTGs by the visual inspection. Galaxy images are drawn from the \textit{HST} in the F814W filter. Recall that our galaxy population resides in a relatively narrow redshift range ($0.2 < z < 0.4$). All images are 5 arcseconds on a side.}
    \label{fig:ltgs}
\end{figure*}

\begin{figure}
    \centering
    \includegraphics[width=0.9\columnwidth]{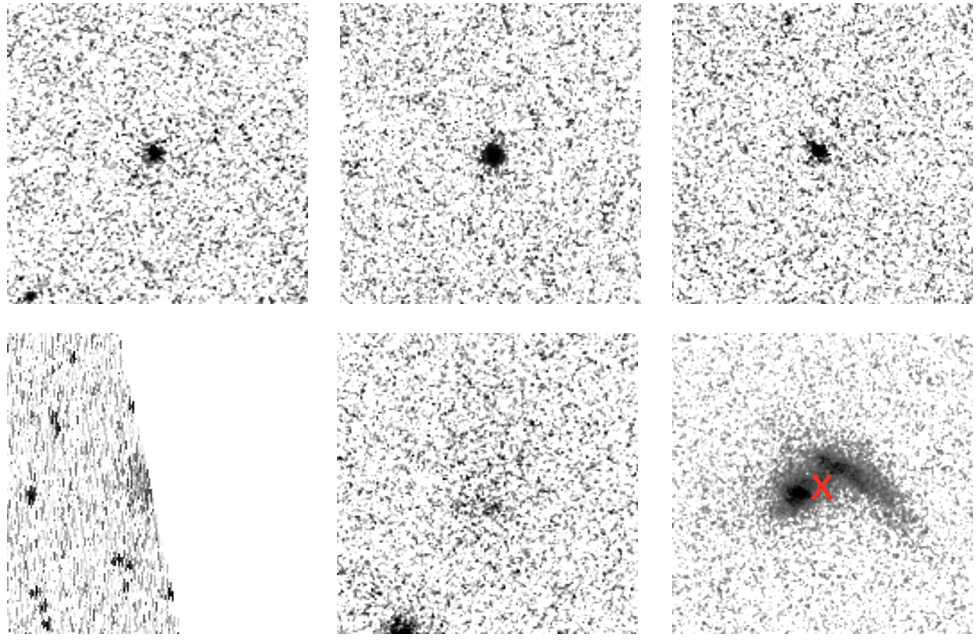}
    \caption{Examples of galaxies classified as compact (top row) by the visual inspection and those that are unclassifiable (bottom row). Galaxy images are taken from the \textit{HST} in the F814W filter. In the bottom row, the first galaxy (from the left) has a blank image and the second does not exhibit enough flux for a classification. The third system contains two galaxies that are blended in the COSMOS2020 catalogue (the COSMOS2020 centroid is shown by the red cross). All images are 5 arcseconds on a side.}
    \label{fig:unc_cmp}
\end{figure}

\begin{figure*}
    \centering
    \includegraphics[width=\textwidth]{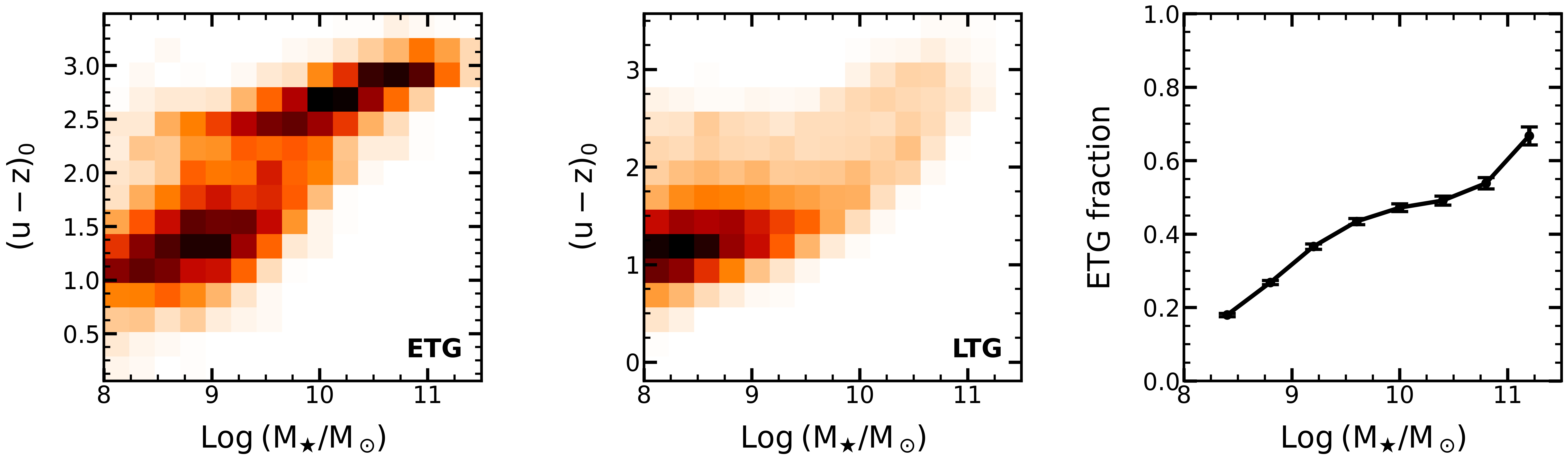}
    \caption{The left and middle panels show the distributions of ETGs and LTGs in the rest-frame $(u-z)$ colour vs stellar mass plane respectively. The right-hand panel shows the ETG fraction as a function of stellar mass. The uncertainties in the right-hand panel are calculated following \citet[][]{Cameron2011}.}.
    \label{fig:mass_colour}
\end{figure*}

Recent work has suggested that the relationship between galaxy properties and environment can be understood more precisely by looking at the details of the \textit{locations} of galaxies in the cosmic web, rather than by just considering the numerical value of the local density. For example, \citet{Laigle2018} demonstrate that the most massive galaxies reside statistically closer to their nearest filament. At fixed stellar mass, passive (i.e. redder) galaxies are also found closer to filaments, regardless of the numerical value of the local density. In a similar vein, \citet{Krajlic2020} show that less star forming (i.e. redder) and less rotation-supported (i.e. more early-type) galaxies tend to be connected to a larger number of filaments. Both of these results appear to be in general agreement with the predictions of cosmological simulations. In other words, the properties of massive galaxies (like stellar mass and colour) appear to be strongly connected with their locations in the cosmic web.  

Despite a great deal of knowledge about how the properties of massive galaxies correlate with environment, a statistical exploration of how the relationship between galaxy morphology and location in the cosmic web changes from the massive to dwarf galaxy regime is currently missing. This is particularly true in relatively low-density environments (which host the majority of galaxies), because, as we discuss below, constructing complete, unbiased samples of dwarfs in such environments has ben challenging using past surveys. Accurately probing the relationship between morphology and environment requires a mass-complete sample of galaxies. In other words, we wish to use a galaxy population in which all objects at a given stellar mass, regardless of their star formation history, exist in the sample. This is because morphology tends to correlate with star formation history, with ETGs tending to host relatively older (and therefore fainter and redder stellar populations) than their late-type counterparts \citep[e.g.][]{Lazar2024a}. Mass-incompleteness skews galaxy populations that appear in a survey towards bluer, late-type objects, which, in turn, may introduce biases in our analysis of how morphology is connected with environment. 

Given their intrinsic faintness, typical dwarfs are not detectable in past large surveys like the SDSS spectroscopic sample \citep{Alam2015}, because they are too shallow. As discussed in detail in \citet{Kaviraj2025}, the consequence of this strong selection effect is that the dwarfs that do exist in datasets like the SDSS host significant levels of star formation activity, which boost the luminosities of these (otherwise undetectable) dwarfs above the detection limits of shallow surveys. Thus, the dwarf populations in shallow surveys like SDSS are not mass-complete and are significantly biased in favour of blue systems, making them unrepresentative of the dwarf population as a whole. While completeness is possible to achieve in the very local Universe, such as in the Local Volume or in relatively high density regions like local clusters, \citealt[e.g.][]{Tolstoy2009,Duc2015,Geha2017,Venhola2018,Trujillo2021,Poulain2021,Mao2021}) these locations either skew the galaxy population towards dense regions or only select dwarfs associated with nearby massive systems (and therefore do not sample a large spectrum of environments). 

Constructing galaxy populations, outside the very local Universe, that are mass-complete down into the dwarf regime, requires surveys that are both deep and wide. This will become routinely possible in the near future using datasets like the Legacy Survey of Space and Time \citep[LSST, e.g.][]{Ivezic2019} and \textit{Euclid} \citep{Euclid2025}. However, deep data with modest footprints and high resolution optical imaging already exists in fields like COSMOS, which enable us to probe the connection between morphology and environment across the massive and dwarf-galaxy regimes, albeit in relatively small areas of the sky. 

In this study we explore open questions about the connection between the \textit{Hubble Space Telescope (\textit{HST})}-derived morphology of galaxies and their location in the cosmic web using a mass-complete sample of $\sim$13,000 galaxies in COSMOS which have stellar masses and redshifts in the ranges 10$^8$ M$_{\odot}$ < $M_{\star}$ < 10$^{11.5}$ M$_{\odot}$ and $0.2<z<0.4$, respectively. For example, how does morphology correlate with local density and galaxy distances to nodes and filaments? Is the correlation stronger in some of these quantities than in others? How do these relationships change as a function of stellar mass from the massive to dwarf galaxy regime? 

The plan for this paper is as follows. In Section \ref{sec:data}, we describe the construction of a mass-complete galaxy sample of $\sim$ 13,000 galaxies in the COSMOS field that underpins this study, the classification of galaxy morphologies via visual inspection and the calculation of environmental parameters (local density and galaxy distances from the nodes and filaments that define the cosmic web). In Section \ref{sec:analysis}, we study the connection between morphology, local density and location in the cosmic web. We summarise our findings in Section \ref{sec:summary}.

\section{Data}
\label{sec:data}

\subsection{Galaxy sample}
Our galaxy sample is taken from the Classic version of the COSMOS2020 catalogue, which provides physical parameters, such as stellar masses, photometric redshifts and rest-frame colours, for $\sim$1.7 million sources in the $\sim$2 deg$^2$ COSMOS \citep{Scoville2007} field (centered at 10h, +02$^{\circ}$). The parameters are calculated using the \textsc{LePhare} SED-fitting algorithm \citep{Arnouts2002,Ilbert2006} by exploiting deep photometry in around 40 broad and medium band filters spanning the UV through to 4.5 $\mu$m. The photometric data is drawn from the following instruments: GALEX \citep{Zamojski2007}, MegaCam/CFHT \citep{Sawicki2019}, ACS/\textit{HST} \citep{Leauthaud2007}, Hyper Suprime-Cam \citep{Aihara2019}, Subaru/Suprime-Cam \citep{Taniguchi2007,Taniguchi2015}, VIRCAM/VISTA \citep{McCracken2012} and IRAC/Spitzer \citep{Ashby2013,Steinhardt2014,Ashby2015,Ashby2018}. 

The COSMOS2020 object detection incorporates optical ($i,z$)  from the UltraDeep layer of the HSC Subaru Strategic Program (HSC-SSP; \citealt{Aihara2019} ), which offers a point-source depth of $\sim$28 mag (around $\sim$10 mag fainter than the magnitude limit of the SDSS spectroscopic main galaxy sample). Optical and infrared aperture photometry are produced using the \textsc{SExtractor} \citep{Bertin1996} and \textsc{IRACLEAN} \citep{Hsieh2012} codes respectively. The SED fitting yields photometric redshift uncertainties that are smaller than $\sim$1 per cent for bright ($i<22.5$ mag) galaxies and smaller than $\sim$4 per cent for their faint ($25<i<27$ mag) counterparts.   

To ensure that our sample is mass-complete, we follow the methodology described in \citet{Kaviraj2025}. We define the completeness redshift for a galaxy population of a given stellar mass as the maximum redshift at which a purely old simple stellar population (SSP), that forms in an instantaneous burst at $z=2$ and has the corresponding stellar mass, is detectable in the HSC UltraDeep imaging (that underpins object detection in COSMOS2020). This purely-old SSP represents a faintest `limiting' case, since real galaxies are not composed solely of old stars and will therefore be more luminous than this limiting value. It follows therefore that, if this limiting case is detectable at the depth of a given survey, then the entire galaxy population (at a given stellar mass) should also be detectable. Figure 2 in \citet{Kaviraj2026} indicates the completeness redshifts, calculated using this method, for various datasets including COSMOS2020 (solid orange line). This figure indicates that the COSMOS2020 galaxy catalogue is likely to be complete down to stellar masses of 10$^{8}$ M$_{\odot}$, out to at least $z\sim0.4$. As a result, we select the parent population of objects in this paper to be COSMOS2020 galaxies that have $M_{\star}$ > 10$^{8}$ M$_{\odot}$ and $z<0.4$. 


\subsection{Galaxy morphologies via visual inspection of \textit{HST} images}
\label{sec:morphologies}

A wide variety of methods have traditionally been used in the observational literature to classify the morphological properties of galaxies (largely focussed on the massive-galaxy regime). These range from direct visual inspection of images \citep[e.g.][]{Kaviraj2014b,Lintott2011} to algorithmic methods which employ morphological parameters \citep[e.g.][]{Conselice2003,Lotz2004,Rodriguez-Gomez2019,Lazar2024a,Sazonova2026} or, more recently, machine learning techniques \citep[e.g.][]{Martin2020,Uzeirbegovic2020,Lazar2023,Walmsley2022}. In the massive regime, properties like rest-frame colour show correlations with galaxy morphology \citep[e.g.][]{Strateva2001} and are sometimes used as a proxy for the latter. However, the morphological purity of such colour-selected samples is relatively low \citep[e.g.][]{Uzeirbegovic2022}. Algorithmic and colour-dependent methods are typically calibrated against visual inspection and are useful for studying extremely large datasets for which visual inspection can be prohibitively time consuming. 

However, recent work has shown that morphological classification using algorithmic methods and those that employ colour is significantly more challenging in dwarf galaxies than in their massive counterparts. For example, galaxies become progressively less concentrated at lower stellar masses \citep[e.g.][]{Lazar2024a,Lazar2024b}. Thus, while ETGs are significantly more concentrated than LTGs in the massive regime,  these differences are much smaller in dwarfs leading to a loss in the leverage that galaxy concentration provides in morphological classification. This, in turn, means that algorithmic methods are less effective at separating morphological classes in dwarfs than in massive galaxies. In a similar vein, while ETGs and LTGs show a large separation in colour in the massive-galaxy regime, the colour distributions of dwarf ETGs and LTGs show significant overlap \citep[e.g.][]{Lazar2024a} making colour selection unviable for identifying morphological types in the dwarf regime. As a result of these challenges, visual inspection, as is the case in this study, becomes essential for the morphological classification of dwarf galaxies.



We visually inspect \textit{HST} F814W images in COSMOS \citep{Koekemoer2007,Massey2010}, which have a point spread function full width at half maximum  of $\sim$0.1 arcseconds, to morphologically classify our galaxies. The visual inspection is carried out by one expert classifier (SK). The images are randomised and galaxy stellar masses and redshifts are kept hidden to avoid introducing any classification biases. 

The visual inspection separates the galaxy population into two broad morphological classes which have traditionally underpinned the literature on the link between morphology and environment. The first are `early-type' galaxies (ETGs) which, by their traditional definition, exhibit strong central light concentrations but otherwise smooth light distributions \citep[e.g.][]{deVaucouleurs1959}. The second are `late-type' galaxies (LTGs) which do not exhibit such central light concentrations and typically show one or more types of structure (e.g. spiral arms, clumps etc., e.g. \citealt{Buta2011}). A small fraction of objects while resolved, are somewhat too small to classify securely and are classified as `compact' and not used in the analysis. These systems cluster towards the lower end of our stellar mass range. While the compact population cannot be considered separately in our analysis due to their low numbers, re-deriving the results presented in the sections below by combining the compact objects with either the ETGs or LTGs does not alter our conclusions. 

Finally, another small fraction of galaxies are labelled as unclassifiable, either because they do not have enough flux in the \textit{HST} image to make classification possible or because two \textit{HST} objects appear very close to each other suggesting that the (ground-based) HSC detection image could be a blend of two objects, which could compromise the COSMOS2020 photometry. If the mass threshold that separates the massive and dwarf-galaxy regimes is defined as $M_{\star}$ = 10$^{9.5}$ M$_{\odot}$, then the fractions of ETGs, LTGs, compact galaxies and unclassified objects are around 0.48, 0.50, 0.007 and 0.006 in the massive-galaxy population and around 0.21, 0.64, 0.08 and 0.06 in the dwarf regime. 

Figures \ref{fig:etgs} and \ref{fig:ltgs} shows example images of our ETGs and LTGs respectively, while Figure \ref{fig:unc_cmp} presents examples of systems that are compact or unclassified. Figure \ref{fig:mass_colour} shows the distribution of ETGs (left-hand panel) and LTGs (middle panel) in the rest-frame $(u-z)$ colour vs stellar mass plane. In agreement with the literature, ETGs tend to be more massive and dominate the red sequence in the massive-galaxy regime \citep[e.g.][]{Strateva2001}. However, the separation in rest-frame colour between ETGs and LTGs becomes weaker in the dwarf regime, as has already been noted in recent work \citep[e.g.][]{Lazar2024a}. While dwarf LTGs do dominate the bluest colours in the colour -- stellar mass space, a significant fraction of dwarf ETGs are also found in the blue cloud. The right-hand panel of this figure shows how the ETG fraction evolves as a function of stellar mass in our galaxy population. The overall ETG fraction in the galaxy population decreases monotonically from $\sim$65 per cent at $M_{\star}$ $\sim$ 10$^{11}$ M$_{\odot}$ to $\sim$20 per cent at $M_{\star}$ $\sim$ 10$^{8}$ M$_{\odot}$.


\subsection{Measuring local density and topological structure}
\label{sec:disperse}

A key aim of this study is to study how galaxy morphology changes, at a given stellar mass, with both local density and location in the cosmic web. To measure these quantities we utilise the structure-finding algorithm DisPerSE \citep{Sousbie2011}, to measure the numerical value of local density and the distances of galaxies to the nearest nodes and filaments in the cosmic web. DisPerSE exploits Delaunay tessellations to measure the local density field, which is calculated using the positions of galaxies \citep{Schaap2000}. It then uses segments to connect topological saddle points with local maxima, forming a set of ridges that describes the network of filaments that define the cosmic web. The minima, maxima and saddles in the density map correspond to the locations of voids, nodes and the centres of filaments respectively. We refer readers to \citet{Sousbie2011} for further details of the algorithm. 

We follow the methodology of \citet{Lazar2023} and \citet{Bichanga2024}, who have recently used DisPerSE to perform a similar density analysis using the COSMOS2020 catalogue. The properties of the filament network created by the algorithm are determined by a `persistence' parameter, which sets a threshold value for defining pairs of critical points within the density map. A persistence of \textit{N} produces a network where all critical pairs with Poisson probabilities below \textit{N}$\sigma$ from the mean are removed. Here we use a persistence of 2, which removes ridges that are close to the noise level and could therefore be spurious. 


The accuracy of the COSMOS2020 redshifts enables us to employ relatively narrow, well-defined redshift slices in which we build our density maps. The maps are constructed using galaxies with $M_{\star}$ > 10$^{10}$ M$_{\odot}$, as these have the smallest redshift errors and dominate the local gravitational potential well. When constructing each map, we weight individual galaxies by the area under their redshift probability density function that is contained within the slice in question. This takes into account the fact that galaxy photometric redshifts, although very accurate in COSMOS2020, have associated errors. For each galaxy we then calculate the local density and the projected distances from the nearest node and filament, which we use for our subsequent analysis. It is worth noting here that \citet{Laigle2018} have shown, using the Horizon-AGN cosmological simulation \citep{Kaviraj2017}, that datasets like COSMOS2020 which offer high photometric redshift precision successfully recover the broad 3D properties of the cosmic web from 2D projected density maps. 

\begin{figure*}
    \centering
    \includegraphics[width=0.8\textwidth]{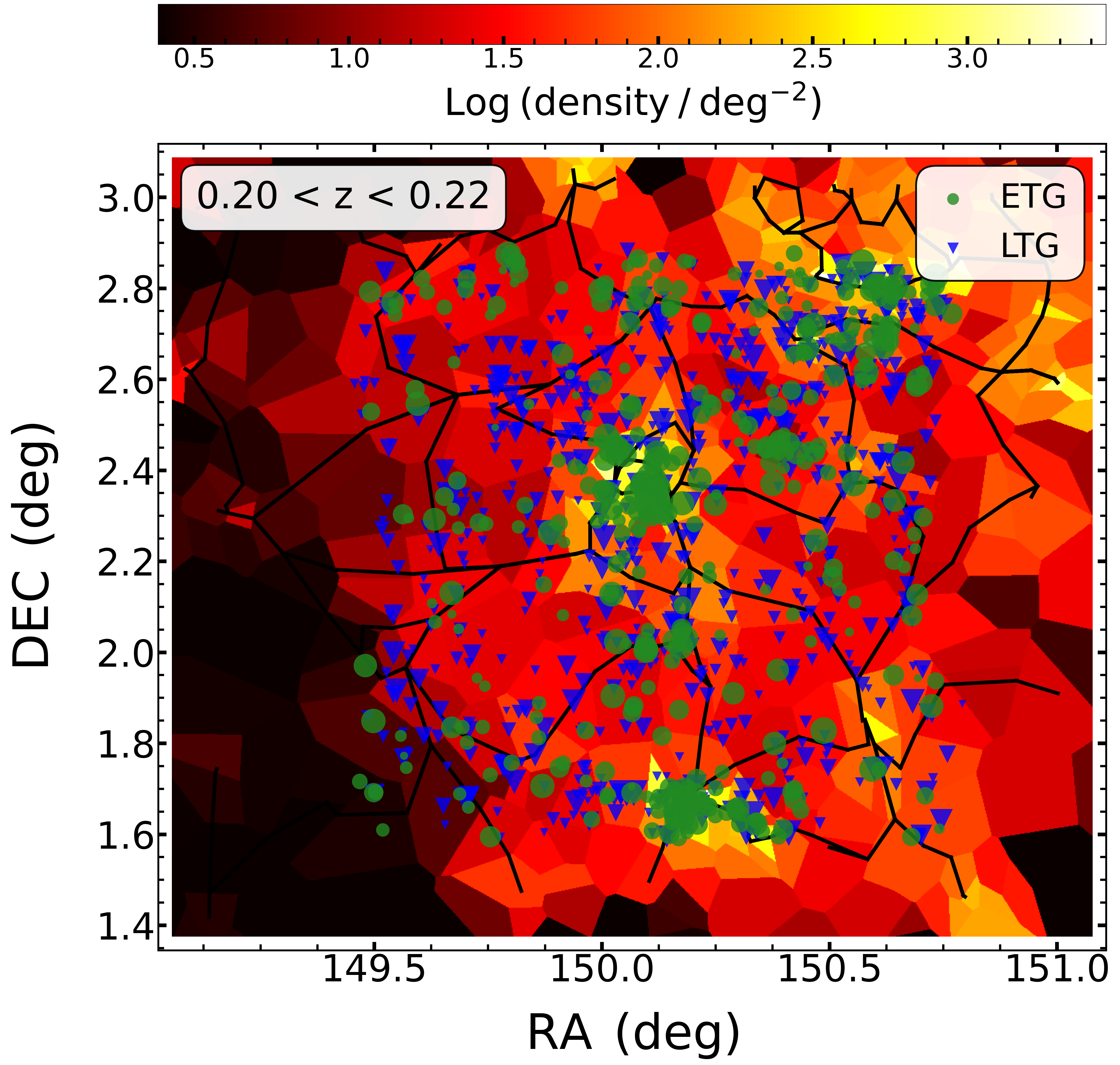}
    \caption{An example density map at $z \sim 0.2$ with ETGs (green) and LTGs (blue) labelled on top. The size of each marker is proportional to the log stellar mass of the galaxy. The smallest marker corresponds to a stellar mass of $M_{\star}$ $=$ $10^{8}$ M$_{\odot}$ and the largest marker corresponds to a stellar mass of $M_{\star}$ $=$ $10^{11.5}$ M$_{\odot}$.}
    \label{fig:example_density_map}
\end{figure*}

Figure \ref{fig:example_density_map} shows an example density map at $z \sim 0.2$ with ETGs and LTGs labelled on top. Note that, since there are more galaxies in the frame at higher redshifts, the density values in the redshift slices slowly drift towards slightly higher values as redshift increases. Hence, in our analysis in Section \ref{sec:analysis} below, we use a `normalised density', where the density values of galaxies in a given slice are divided by the 50th percentile value (i.e. the median value) of the density distribution within that slice. Finally, as discussed in \citet{Kaviraj2025}, the small footprint of the COSMOS field (2 deg$^2$) means that the number density of massive galaxies at $z<0.2$ is small and not sufficient for constructing density maps. Our study is therefore based on the $\sim$ 13,000 COSMOS2020 galaxies which have stellar masses in the range $M_{\star}$ > 10$^{8}$ M$_{\odot}$ and redshifts in the range $0.2<z<0.4$.

\begin{figure}
    \centering
    \includegraphics[width=\columnwidth]{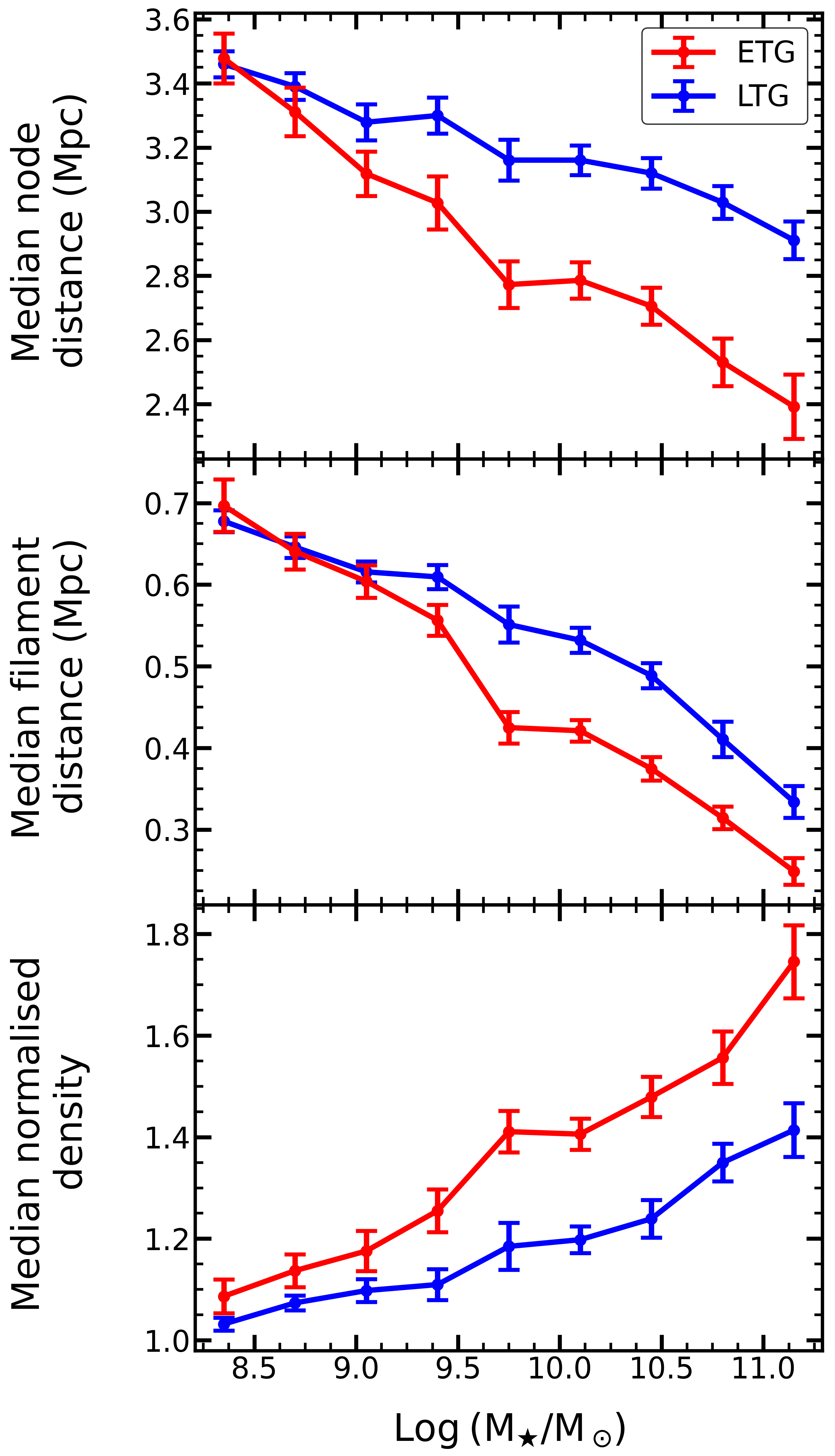}
    \caption{Median projected distances to nodes (top) and filaments (middle) and the median normalised densities (bottom) of ETGs (red) and LTGs (blue) as a function of stellar mass. The errors on the median are calculated via bootstrapping.}
    \label{fig:median_distancesANDdens}
\end{figure}

\begin{figure}
    \centering
    \includegraphics[width=\columnwidth]{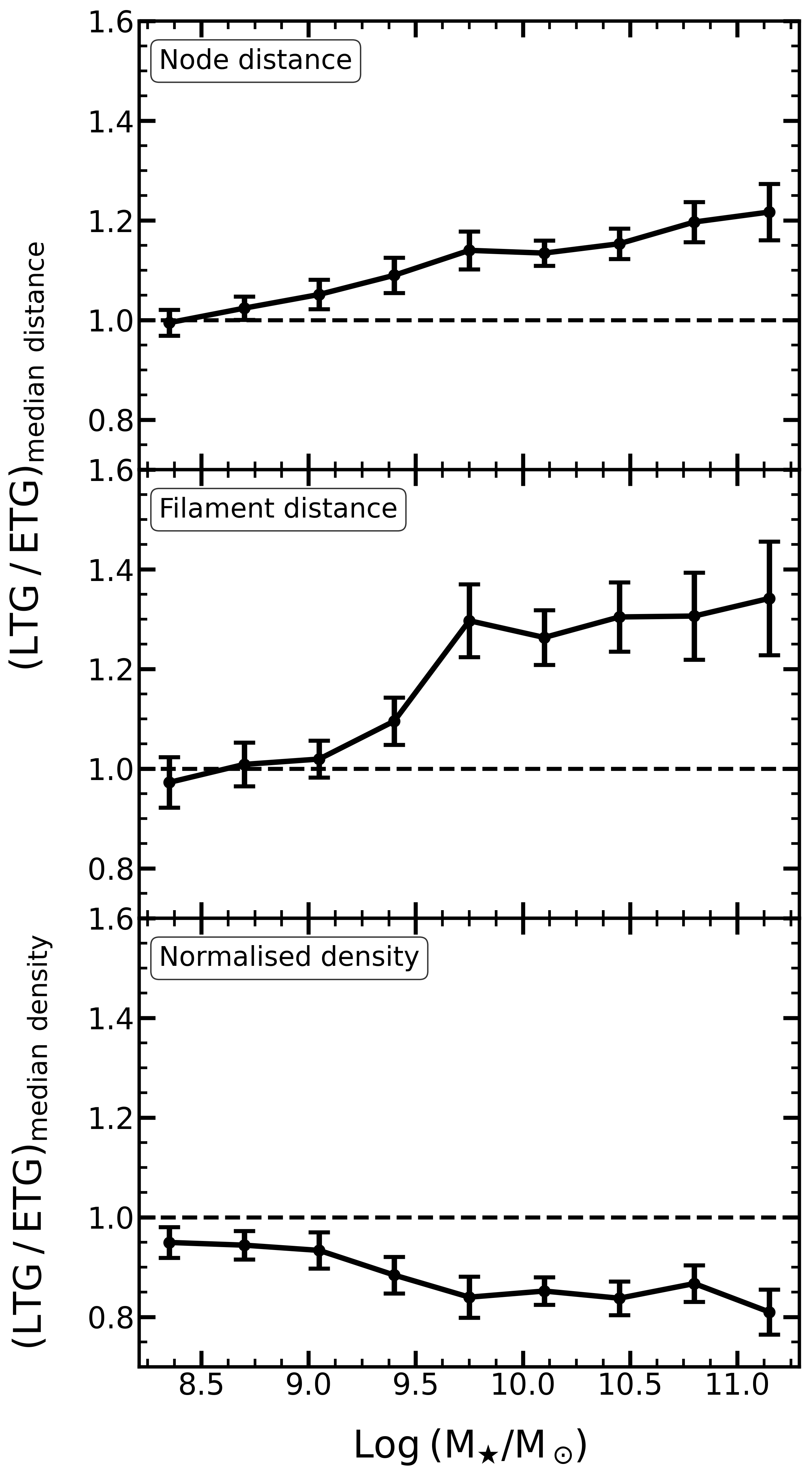}
    \caption{The ratio of the median LTG and ETG distances to nodes (top panel) and to filaments (middle panel). The bottom panel shows the ratio of the LTG and ETG median normalised densities. The dashed horizontal line corresponds to a ratio of 1. The errors on the ratios are calculated using standard error propagation.}
    \label{fig:ratio_median_distancesANDdens}
\end{figure}

\begin{figure*}
    \centering
    \includegraphics[width=\textwidth]{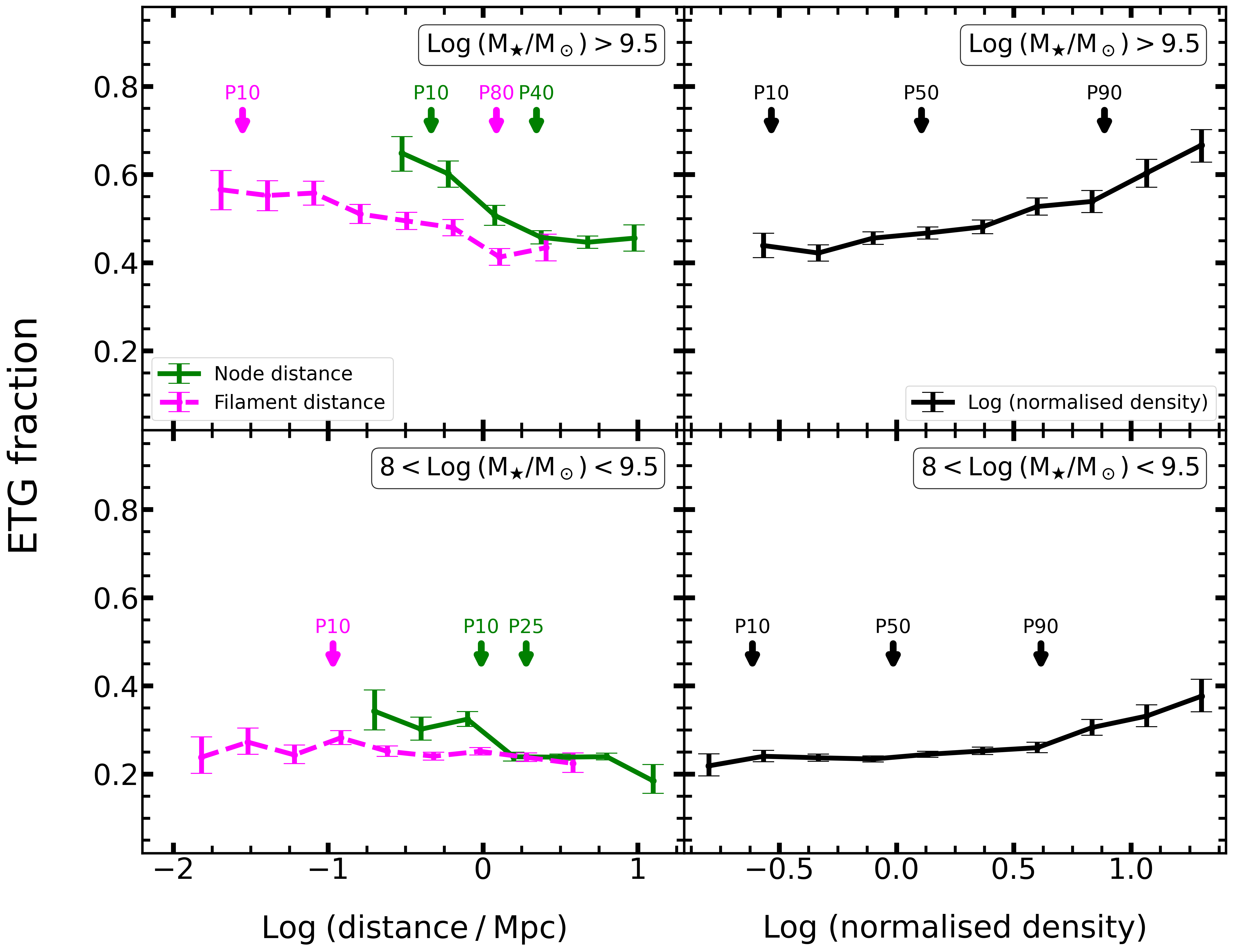}
    \caption{The left-hand column shows the ETG fraction for massive galaxies (top) and dwarf galaxies (bottom) as a function of the distances to nodes (green) and filaments (magenta). The right-hand column shows the ETG fraction as a function of the log of the normalised density for the massive galaxies (top) and dwarf galaxies (bottom). The arrows annotated with `Px' indicate the `x' percentile value (e.g. `P10' indicates the 10$^{\rm th}$ percentile value and so on). The errors on the ETG fractions are calculated following \citet{Cameron2011}.}
    \label{fig:ETG_fractions}
\end{figure*}

It is instructive to consider the types of large-scale structures that are likely to exist within our COSMOS2020 footprint. We note first that the COSMOS field is not centred on a known region of high density. We consider the \textit{M}$_{200}$ values, which are proxies for the halo virial masses, of groups identified in the literature \citep{Finoguenov2007,George2011,Gozaliasl2014,Gozaliasl2019} in COSMOS at our redshifts of interest ($0.2<z<0.4$) These virial masses lie in the range 10$^{12.9}$ M$_{\odot}$ < \textit{M}$_{\rm{200}}$ < 10$^{13.8}$ M$_{\odot}$, with a median value of 10$^{13.4}$ M$_{\odot}$. For comparison, a small cluster like Fornax has a virial mass of $\sim$10$^{13.9}$ M$_{\odot}$ \citep{Drinkwater2001}, while large clusters like Virgo and Coma have virial masses of $\sim$10$^{15}$ M$_{\odot}$ \citep[e.g.][]{Fouque2001,Gavazzi2009}. The galaxy population considered in this study therefore spans a wide spectrum of environments from large groups to the field in agreement with the conclusions of other recent work \citep[e.g.][]{Martin2025}. We note, however, that our galaxies do not reside in rich clusters. 


\section{The relationship between galaxy morphology and location in the cosmic web}
\label{sec:analysis}

We begin, in Figure \ref{fig:median_distancesANDdens}, by showing the median (projected) distances to nodes (top) and filaments (middle) of ETGs and LTGs as a function of stellar mass. While ETGs typically reside closer to nodes, the differences between ETGs and LTGs in their distances to the nearest node diminish with stellar mass, until they become negligible around $M_{\star}$ $\sim$ $10^{8.5}$ M$_{\odot}$. A similar pattern is seen in the distances of ETGs and LTGs to their nearest filament, with the differences becoming negligible at a slightly larger stellar mass of around $M_{\star}$ $\sim$ $10^{9}$ M$_{\odot}$. In other words, in the massive-galaxy regime, ETGs tend to reside closer to the cosmic web than LTGs. The differences in the distance to the cosmic web between ETGs and LTGs become progressively smaller with decreasing stellar mass, until they disappear in the dwarf regime around $M_{\star}$ $\sim$ $10^{9}$ M$_{\odot}$. The bottom panel of Figure \ref{fig:median_distancesANDdens} shows the median values of the normalised density of ETGs and LTGs as a function of galaxy stellar mass. Not unexpectedly, ETGs tend to inhabit denser regions than their LTG counterparts at all stellar masses. Recall that the normalised density is defined as the density divided by the median density in the redshift slice in question. 

In Figure \ref{fig:ratio_median_distancesANDdens}, we summarise the relative differences between ETGs and LTGs by studying the ratios (LTG/ETG) of the median values of different environmental parameters as a function of stellar mass. The top and middle panels present the ratio of the median distances of LTGs and ETGs to nodes and filaments respectively, as a function of stellar mass. A ratio of 1 indicates no difference, while a value larger than 1 indicates that LTGs lie further away from nodes or filaments. The distances of LTGs from nodes (filaments) are a factor of 1.2 (1.4) larger at the upper end of the stellar mass range probed by this study ($M_{\star}$ $\sim$ $10^{11}$ M$_{\odot}$). As would be expected from Figure \ref{fig:median_distancesANDdens}, these ratios decrease with decreasing stellar mass and become consistent with 1 at the lower end of our stellar mass range ($M_{\star}$ $\sim$ $10^{8.5}$ M$_{\odot}$). The bottom panel of this figure indicates that the ratio between the local densities of LTGs and ETGs is less than 1 (i.e. LTGs reside in less dense regions), mirroring the findings in the bottom panel of Figure \ref{fig:median_distancesANDdens}.

We proceed, in Figure \ref{fig:ETG_fractions}, by exploring how the morphological mix of galaxies is connected to environment. We quantify the morphological mix using the fraction of galaxies that are ETGs. In these plots, we indicate the positions of some indicative percentile values, using the vertical arrows (e.g. `P10' refers to the 10th percentile value).The left-hand column of this figure shows that the ETG fraction rises towards smaller values of distances to both nodes and filaments. The effect is stronger in massive galaxies ($M_{\star}$ > $10^{9.5}$ M$_{\odot}$; top panel) than in their dwarf counterparts ($10^{8}$ M$_{\odot}$ < $M_{\star}$ < $10^{9.5}$ M$_{\odot}$; bottom panel). This is expected, given the analysis above, since the differences between ETGs and LTGs, in terms of their distances to the cosmic web, diminish with decreasing stellar mass. In both massive and dwarf galaxies, the rise in the ETG fraction is more significant with decreasing values of the node distance than the filament distance. 


The right-hand column of this figure shows the corresponding plots as a function of the normalised density. We find that, while the ETG fraction in the massive regime increases steadily with increasing (log) normalised density, it only shows a mild evolution with density in the dwarf regime until around the 90$^{\rm th}$ percentile of the normalised density distribution where it starts to increase appreciably. Recall that, in a similar vein, the ETG fraction in the dwarf regime shows a significant rise only for the smallest 10 percentile values of the node distance distribution (lower left panel of this figure) and does not show any change with decreasing filament distance. Taken together, this suggests that morphological transformation is likely to be more tightly correlated with environment in the massive than the dwarf regime. Thus, in the massive regime, increases in local density are always reflected in corresponding changes in the ETG fraction. However, the ETG fraction in the dwarf regime responds strongly to increasing local density only when the density values are in the upper 10 percentiles of the density distribution (which typically correspond to regions close to nodes).  


Our findings, at least in the massive-galaxy regime, appear aligned with studies that have explored various galaxy properties as a function of both local density and location in the cosmic web. For example, \citet{Laigle2018} demonstrate that in the $M_{\star}$ > $10^{10}$ M$_{\odot}$ regime, at fixed stellar mass, passive objects reside closer to the cores of filaments than their star forming counterparts. \citet{Kraljic2018} use massive galaxies from the GAMA survey \citep{Driver2009} to show that galaxies that are more massive and/or quiescent are typically located closer to filaments than their less massive and/or star-forming counterparts. The galaxy red fraction also increases closer to nodes (and also to filaments when the distance to nodes is controlled for). Similarly, \citet{Krajlic2020} demonstrate that, for nearby massive galaxies in the SDSS, being more massive, less star-forming and more pressure-supported (i.e. more early-type) increases the likelihood of residing closer to nodes that have higher `connectivity' i.e. are connected to a greater number of filaments. \citet{Malavasi2017} confirm that this tendency of more massive or quiescent galaxies to reside closer to the cosmic web persists out to at least $z\sim 0.7$ (see also \citealt{Poudel2017}). 

While the aforementioned studies do not probe the dwarf regime, it is interesting to note that \citet{Laigle2018} do find that the difference between passive and star-forming galaxies in the massive regime, in terms of their distances from the cosmic web, diminishes with decreasing stellar mass. This appears consistent with the results of \citet[][]{Nandi2026}, who find that the differences between the galaxy red fractions in different environments (e.g. sheets, filaments and clusters) diminish for objects at the lower end of the stellar mass regime they consider ($10^{10}$ M$_{\odot}$ < $M_{\star}$ < $10^{10.6}$ M$_{\odot}$). Similarly, \citet{Hoosain2024}, who also employ DisPerSE to trace the cosmic web in the very nearby Universe ($z<0.02$), show that both the red and gas-poor fractions increase with decreasing distance from the cosmic web until $M_{\star}$ $\sim$ $10^{9}$ M$_{\odot}$, below which this segregation disappears. Since ETGs and LTGs tend to have a preference for passive and star-forming systems (although it is important to note that this mapping is not perfect, see e.g. \citealt{Lazar2024a}), these works appear to find qualitatively similar trends as those in our study i.e. that the differences between ETGs and LTGs, in terms of their locations in the cosmic web, diminish as stellar mass decreases. 



We suggest an explanation for our findings as follows. Figure \ref{fig:median_distancesANDdens} indicates that lower mass galaxies reside further away from the core regions of filaments. Since the filaments themselves have a finite extent, this suggests that lower mass galaxies, regardless of morphological type, are likely confined to a smaller region of the filament itself. This would explain why the differences in the distances between the ETGs and LTGs tend to diminish with decreasing stellar mass. 

The physical processes that operate in different locations within the cosmic web are also different. Greater proximity to nodes likely inhibits coherent angular momentum acquisition, while residing in the core regions of filaments increases the impact of mergers and interactions. Both these locations are likely to increase the probability of galaxies becoming less rotationally-supported (i.e. more ETG like). This implies that, since more massive galaxies reside closer to the nodes and the core regions of filaments, the probability of massive galaxies becoming ETGs is higher. Given that the ETG fraction can be considered to be a proxy for this probability, this explains why the ETG fraction increases with stellar mass, (see right-hand panel of Figure \ref{fig:mass_colour}). It also explains why, in the massive-galaxy regime, ETGs are found closer to nodes and filaments than their LTG counterparts. 

It is worth considering our results in unison with recent work which has studied other aspects of galaxy evolution across the massive and dwarf-galaxy regimes. For example, \citet{Lazar2026} have demonstrated that the concept of downsizing, i.e. progressively lower-mass galaxies maintaining their star formation until later epochs, is incorrect. Instead of monotonically decreasing with decreasing stellar mass, the red/quenched galaxy fraction shows a U-shape in the dwarf regime and actually begins to rise again as the galaxy stellar mass falls below $\sim$10$^{8.5}$ M$_{\odot}$. This U-shape persists in different environments (albeit with a higher normalisation in denser environments). \citet{Lazar2026} show that the shape of the red fraction is consistent with supernova feedback dominating quenching at $M_{\star}$ < 10$^{8.5}$ M$_{\odot}$ and a mix of supernova and AGN feedback controlling the quenching at $M_{\star}$ > 10$^{8.5}$ M$_{\odot}$. The lack of dependence of the U-shaped red fraction with environment in dwarf galaxies suggests that, in this regime, the quenching process is principally driven by internal processes and not by environment. This mirrors our result which suggests that the process of ETG production, at least in groups and the field, does not have a strong connection to local environment. 

Finally, we note that \textit{how} dwarf ETGs actually form in the (relatively low density) environments studied here, remains an open question. Our work indicates that internal processes like supernova feedback, augmented by the influence of AGN, which have recently been shown to be common in dwarfs \citep[e.g.][]{Kaviraj2017,Koudmani2021,Mezcua2024,Kaviraj2026} could have an important role in determining the stellar content and morphological properties of dwarfs. Taken together, this suggests that many aspects of galaxy evolution are increasingly driven by internal processes rather than environment as stellar mass decreases.

\section{Summary}
\label{sec:summary}

We have explored the connection between galaxy morphology and location in the cosmic web using a mass-complete sample of $\sim$13,000 galaxies, 
in the stellar-mass and redshift ranges 10$^8$ M$_{\odot}$ < $M_{\star}$ < 10$^{11.5}$ M$_{\odot}$ and $0.2<z<0.4$, respectively. Our galaxy sample resides in relatively low-density environments which host the majority of galaxies (groups and the field) and does not inhabit dense regions like clusters. We have split our galaxy population into ETGs and LTGs via visual morphological classifications of \textit{HST} images, calculated local density and galaxy distances from nodes and filaments using the DisPerSE algorithm and studied how the environments of ETGs and LTGs behaves as a function of stellar mass. Our main conclusions are as follows: 

\begin{itemize}

    \item The fraction of galaxies that are ETGs shows a strong monotonic decrease with decreasing stellar mass, from $\sim$60 per cent at $M_{\star}$ $\sim$ 10$^{11}$ M$_{\odot}$ to $\sim$20 per cent at $M_{\star}$ $\sim$ 10$^{8}$ M$_{\odot}$. 

    \item Regardless of morphological type, lower mass galaxies typically reside further away from nodes and the core regions of filaments than their more massive counterparts. 

    \item Additionally, in the mass range $M_{\star}$ $\gtrsim$ 10$^{9}$ M$_{\odot}$, ETGs reside at smaller distances from nodes and filaments than LTGs, although this environmental segregation becomes progressively weaker as stellar mass decreases. Below $M_{\star}$ $\sim$ 10$^{9}$ M$_{\odot}$ ETG and LTGs are found at similar distances from the cosmic web.
    
    \item The diminishing differences between ETGs and LTGs at lower stellar mass are likely due to the fact that filaments have a finite extent. Since lower mass galaxies, regardless of  morphological type, lie further away from filament cores, they are likely confined to a smaller region of the filament, which explains the lack of strong locational differences between ETGs and LTGs. 

    \item For high stellar mass galaxies (where ETGs and LTGs show strong environmental segregation), greater proximity to nodes likely inhibits coherent angular momentum acquisition, while residing closer to filament cores increases the likelihood of interactions. Both of these factors increase the probability of creating dispersion-dominated systems, resulting in a sharp rise of the ETG fraction close to nodes (and, to a lesser extent, filaments). This also explains the decreasing ETG fraction at lower stellar mass, since dwarf galaxies are much less likely to reside close to nodes and filament cores, where the processes that create dispersion-dominated objects are likely to operate.  

    \item Taken together with the recent literature, our results suggest that, as stellar mass decreases, both the star formation and structural evolution of galaxies is increasingly dominated by internal processes, rather than environment. 
    
\end{itemize}


\section*{Acknowledgements}

SK, IL and AEW acknowledge support from the STFC (grant numbers ST/Y001257/1 and ST/X001318/1). SK also acknowledges a Senior Research Fellowship from Worcester College Oxford. 

The Hyper Suprime-Cam (HSC) collaboration includes the astronomical communities of Japan and Taiwan, and Princeton University. The HSC instrumentation and software were developed by the National Astronomical Observatory of Japan (NAOJ), the Kavli Institute for the Physics and Mathematics of the Universe (Kavli IPMU), the University of Tokyo, the High Energy Accelerator Research Organization (KEK), the Academia Sinica Institute for Astronomy and Astrophysics in Taiwan (ASIAA), and Princeton University. Funding was contributed by the FIRST program from the Japanese Cabinet Office, the Ministry of Education, Culture, Sports, Science and Technology (MEXT), the Japan Society for the Promotion of Science (JSPS), Japan Science and Technology Agency (JST), the Toray Science Foundation, NAOJ, Kavli IPMU, KEK, ASIAA, and Princeton University. This paper makes use of software developed for Vera C. Rubin Observatory. We thank the Rubin Observatory for making their code available as free software at http://pipelines.lsst.io/.

This paper is based on data collected at the Subaru Telescope and retrieved from the HSC data archive system, which is operated by the Subaru Telescope and Astronomy Data Center (ADC) at NAOJ. Data analysis was in part carried out with the cooperation of Center for Computational Astrophysics (CfCA), NAOJ. We are honored and grateful for the opportunity of observing the Universe from Maunakea, which has the cultural, historical and natural significance in Hawaii. This paper used data that is based on observations collected at the European Southern Observatory under
ESO programme ID 179.A-2005 and on data products produced by CALET and
the Cambridge Astronomy Survey Unit on behalf of the UltraVISTA consortium.


\section*{Data availability}

The observational data used in this study are available from \citet{Weaver2022}. The density maps are produced using the DisPerSE algorithm, which is described in \citet{Sousbie2011}.


\bibliographystyle{mnras}
\bibliography{references}



\appendix

\bsp
\label{lastpage}
\end{document}